%% file: main.tex
\pdfoutput=1
\documentclass[sigconf]{acmart}
\usepackage{booktabs} % For formal tables
\usepackage{multirow}
\usepackage{subcaption}
\usepackage{caption}
\usepackage{setspace}
\usepackage{amsmath}
\usepackage[english]{babel}
\definecolor{mygray}{gray}{0.5}
\usepackage{float}
\usepackage{flushend}
\usepackage{mathtools}

\definecolor{midnightgreen}{rgb}{0.0, 0.29, 0.33}
\definecolor{orange}{RGB}{255,127,0}

\begin{document}

\copyrightyear{2019} 
\acmYear{2019} 
\setcopyright{acmlicensed}
\acmConference[SIGIR '19]{Proceedings of the 42nd International ACM SIGIR Conference on Research and Development in Information Retrieval}{July 21--25, 2019}{Paris, France}
\acmBooktitle{Proceedings of the 42nd International ACM SIGIR Conference on Research and Development in Information Retrieval (SIGIR '19), July 21--25, 2019, Paris, France}
\acmPrice{15.00}
\acmDOI{10.1145/3331184.3331198}
\acmISBN{978-1-4503-6172-9/19/07}

\title{Generic Intent Representation in Web Search}

\author{Hongfei Zhang, Xia Song, Chenyan Xiong, Corby Rosset, \\ Paul N.\ Bennett, Nick Craswell, and Saurabh Tiwary}
\affiliation{Microsoft AI \& Research} 
\affiliation{\texttt{\{honzha, xiaso, cxiong, corosset, pauben, nickcr, satiwary\}@microsoft.com}}


\input{Abs.tex}

\maketitle

\input{Introduction.tex}

\input{Related.tex}

\input{Model.tex}

\input{Experiment.tex}

\input{Evaluation.tex}

\input{Applications.tex}

\input{Conclusion.tex}

\section{ACKNOWLEDGEMENT}
We want to thank Tong Wang, Subhojit Som, Maria Kang, Kamara Benjamin and Marc Rapaport for helping with ANN search experiments and human evaluation. We also thank Guoqing Zheng and Susan Dumais for providing valuable feedback for this paper.

\newpage

\bibliographystyle{ACM-Reference-Format}
\normalsize
\bibliography{citation}
% \flushend 
\end{document}

%% file: Abs.tex
\pdfoutput=1
\begin{abstract}
This paper presents GEneric iNtent Encoder (\texttt{GEN Encoder}) which learns a distributed representation space for user intent in search. Leveraging large scale user clicks from Bing search logs as weak supervision of user intent, \texttt{GEN} \texttt{Encoder} learns to map queries with shared clicks into similar embeddings end-to-end and then fine-tunes on multiple paraphrase tasks. Experimental results on an intrinsic evaluation task -- query intent similarity modeling -- demonstrate \texttt{GEN} \texttt{Encoder}'s robust and significant advantages over previous representation methods. Ablation studies reveal the crucial role of learning from implicit user feedback in representing user intent and the contributions of multi-task learning in representation generality. We also demonstrate that \texttt{GEN} \texttt{Encoder} alleviates the sparsity of tail search traffic and cuts down half of the unseen queries by using an efficient approximate nearest neighbor search to effectively identify previous queries with the same search intent. Finally, we demonstrate distances between \texttt{GEN} encodings reflect certain information seeking behaviors in search sessions.
\end{abstract}

%% file: Introduction.tex
\pdfoutput=1
\section{Introduction}

User intent understanding plays a fundamental role in modern information retrieval.
A better understanding of "what a user wants" helps search engines return more relevant documents, suggest more useful queries, and provide more precise answers.
Intent understanding is also challenging: two queries with the same intent may have no term overlap (e.g. ``cheap cars'', ``low-priced autos'') while two queries with slight variations may have completely different meanings (e.g. ``horse racing'', ``racing horses''). A brittle bag-of-words style of query representation faces such challenges as vocabulary mismatch and ambiguity~\cite{croft2010search}.

Distributed representations (embeddings) provide a path to address these challenges and have been increasingly explored with the rapid development of neural methods.
Representing text as continuous embeddings enables \emph{softer matches} than term overlap.
The embeddings also incorporate \emph{additional information}. For example, word2vec maps semantically similar words together by learning they often appear in similar surrounding texts~\cite{word2vec}.
The embeddings also support \emph{richer compositionality} than bag-of-words using neural networks. For example, ELMo and BERT use deep neural architectures to provide more context-aware embeddings~\cite{PetersELMO, BERT}.

Nevertheless, embeddings learned from surrounding texts do not necessarily capture user intents in search queries.
The query ``Harvard student housing'' has similar word2vec embeddings with ``Cornell student housing'',
but its intent --- what a user would find relevant in the search results --- is closer to  ``Cambridge dorm MA''~\cite{K-NRM}. 
Previous work in neural information retrieval demonstrated that these semantic-oriented embeddings often cause topic drift in search~\cite{rekabsaz2017word, zamani2017relevance}. It is often necessary to train task-specific embeddings~\cite{K-NRM, pang2017deeprank, Zamani2018NeuralRM}, while it is not quit clear how to
construct generic representations for search intents.

This work presents \texttt{GE}neric i\texttt{N}tent (\texttt{GEN}) \texttt{Encoder}, which learns a distributed representation space for user intent from user feedback in search. 
\texttt{GEN} \texttt{Encoder} uses a character aware recurrent architecture and a two phase learning strategy.
It first utilizes user clicks as weak supervision of search intent and learns to encode queries which lead to clicks on the same documents (co-click queries) near each other in its representation space. 
Click signals are available at scale in Bing search logs and thus provide sufficient signals to learn \texttt{GEN} \texttt{Encoder} end-to-end.
Then, \texttt{GEN} \texttt{Encoder} employs multi-task learning on paraphrase classification tasks to improve its generality.

We present an intrinsic evaluation -- query intent similarity modeling -- to evaluate the quality of intent representations by their ability to group queries with similar intents together.
Three datasets at different difficulties were constructed with query pairs sampled from search sessions and assessor labels of their intent similarities.
The results demonstrate \texttt{GEN} \texttt{Encoder}'s robust advantages over previous representation methods including, discrete bag-of-words, embedding-based bag-of-words, relevance-based embeddings, and generic encoders designed for modeling language semantics.

Thorough ablation studies reveal the critical role of user clicks in learning intent representations. Simple encoder architectures trained from co-clicks nearly achieve the performances of previous state-of-the-art text encoders, while complex encoder architectures without co-click signals fail to outperform TF-IDF. Thus, 
the source of supervision signals are a more important first consideration than neural architecture.
Results also demonstrate the importance of multi-task learning in building a generic representation space and the effectiveness of the \texttt{GEN} \texttt{Encoder} architecture.

\texttt{GEN} \texttt{Encoder} has many applications in search. 
We first demonstrate that it alleviates the sparsity of tail search traffic. 
Search intelligence highly depends on learning from previous observations of user behaviors on the query~\cite{agichtein2006improving, li2008learning, huang2013learning}.
However, the long-tail nature of search means many queries are rarely, if ever, observed.
\texttt{GEN} \texttt{Encoder} reduces this sparsity by identifying previously observed queries with the same search intent, using approximate nearest neighbor (ANN) search in the continuous space.
Our experiments with an ANN index of 700 million queries demonstrate that this can be achieved with high coverage, accuracy, and practical latency:
ANN search returns neighbors for the majority of queries and finds more than one neighbor with the exact same intent, at the speed of 10 milliseconds per query.
Incorporating the evidence from these semantically similar queries reduces the fraction of unseen queries from $38\%$ to $19\%$, cutting down half of the long tail sparsity.

Finally, we demonstrate emergent behavior where distance in \texttt{GEN} encodings separates common types of query reformulation behavior in a search session: topic change, exploration, specification, and paraphrase. 
Our quantitative study found that the distances between the \texttt{GEN} encoding of query reformulations are nicely separated to a bi-modal distribution, with different ranges corresponding to different information-seeking relationships between the queries.
While not designed explicitly for this purpose, the evidence suggests that \texttt{GEN} \texttt{Encoder} may have future applications in analyzing and understanding user behaviors in sessions.

A generic distributed representation space capturing user intent---plus efficient approximate nearest neighbor retrieval at scale---has the potential to fundamentally change the way search is conducted. 
\texttt{GEN} \texttt{Encoder} is publicly available to facilitate more future exploration in this direction\footnote{\url{Aka.ms/GenEncoder}}.

%% file: Related.tex
\pdfoutput=1
\section{Related Work}

In user intent analysis, Broder's influential study grouped intents (``the need behind the query'') into predefined categories: Informational, Navigational, and Transactional, based on search tasks~\cite{broder2002taxonomy}.
Another dimension in modeling intent is to categorize them into predefined or automatically constructed taxonomies~\cite{broder2007robust, yin2010building}.
The classification is often done by machine learning models, using information from Wikipedia~\cite{hu2009understanding}, query-click bipartite graph~\cite{li2008learning}, and pseudo relevance feedback~\cite{shen2006building}. 
Query intent classification is a standard component in search engines and 
plays a crucial role in vertical search~\cite{li2008learning} and sponsored search~\cite{broder2007robust}.

Many approaches have been developed to improve the bag-of-words query representation.
The weights of query terms can be better estimated by inverse document frequency (IDF)~\cite{croft2010search} and signals from query logs~\cite{Bendersky2011ParameterizedCW}.
The sequential dependency model~\cite{metzler2005markov} incorporated n-grams. 
Query expansion techniques enrich queries with related terms from relevance feedback~\cite{salton1990improving}, pseudo relevance feedback (PRF)~\cite{lavrenko2001relevance}, or external knowledge graphs~\cite{xiong2015fbexpansion}.
These techniques are the building blocks of modern retrieval systems.

Embedding techniques provide many new opportunities to represent queries and model user intent in a distributed representation space.
Zheng and Callan utilize the semantics of word embeddings in query term weighting~\cite{zheng2015}.
Zamani and Croft use word embeddings to build a smooth language model~\cite{zamani2016embedding} and derived the theoretical framework to combine word embeddings into query embeddings~\cite{zamani2016estimating}.
Guo et al. incorporate the soft match between query and document in their word embedding for ad hoc ranking~\cite{guo2016deep}.

Later, it was found that word embeddings trained from surrounding contexts are not as effective in modeling user intent, information needs, or relevancy~\cite{K-NRM, zamani2017relevance}. Word2vec embeddings align word pairs reflecting different user intents together~\cite{K-NRM} and cause topic drift~\cite{rekabsaz2017word}.
It is preferred to learn a distributed representation directly from information retrieval tasks.
Nalisnick et al. compared the effectiveness of word embeddings trained on search queries versus trained on documents in ad hoc retrieval~\cite{nalisnick2016improving}.
Diaz et al. demonstrated that word embeddings trained on each query's PRF documents are more effective in query expansion~\cite{diaz2016query}.
Hamed and Croft trained relevance-base word embeddings from query and PRF term pairs, which are more effective in embedding-based language model~\cite{zamani2017relevance}.
In ad hoc ranking, neural models are more effective if using embeddings learned end-to-end from relevance labels than using embeddings trained from surrounding contexts, showing the different demands of embeddings in search~\cite{K-NRM, dai2018convolutional, Zamani2018NeuralRM, pang2017deeprank}.

Effective as they are, end-to-end learned embeddings are not always feasible: many tasks do not have sufficient training data and not every task is supervised.
This motivated the development of \emph{universal} \emph{representations}, which aims to provide meaningful and generalizable distributed representations for many language processing tasks.
The Universal Sentence Encoder learns a universally effective encoder using multi-task learning on many language modeling tasks~\cite{cer2018universal}.
Contextual embeddings, such as ELMo~\cite{PetersELMO} and BERT~\cite{BERT}, learn deep neural networks from large corpora and provide contextualized embeddings on sentences and longer texts.
These universal representations effectively capture general language semantics and have shown promising results on many language processing tasks, though their impact on search intent representation is unclear.

This paper brings in recent neural techniques to model search queries ($\mathsection$\ref{sec:archi}) and is the first to leverage co-click weak supervisions and multi-task learning to learn distributed intent representations ($\mathsection$\ref{sec:learning}).
We create the first intrinsic evaluation for query intent representations with scaled intent similarity judgments ($\mathsection$\ref{sec:exp}). 
Evaluation results and analyses demonstrate the effectiveness of our method and the contributions of learning from user clicks, multi-tasks, and the encoder architecture ($\mathsection$\ref{sec:eva}).
To illustrate the general impact in search, we demonstrate how \texttt{GEN} \texttt{Encoder} alleviates the sparsity in tail traffic ($\mathsection$\ref{sec:ann}), and how it helps understanding information seeking behaviors in sessions ($\mathsection$\ref{sec:dist})

%% file: Model.tex
\pdfoutput=1
\input{figure/GENDiagram.tex}
\section{Generic Intent Encoder}
\texttt{GEN} \texttt{Encoder} aims to learn a generic representation space that captures user intent in web search.
This section first describes its architecture and then its learning strategy.

\subsection{Architecture} \label{sec:archi}
As shown in Figure~\ref{fig:gendiagram}, \texttt{GEN} \texttt{Encoder} embeds a query $q$ to a continuous vector $\vec{\texttt{GEN}(q)}$ through three components:
a \emph{word embedding} that maps query words to continuous vectors (\ref{fig:diagramembedding}),
a \emph{character-aware embedding} that models rare words with morphology (\ref{fig:diagramembedding}),
a \emph{mix encoder} that composes word embeddings to query encoding (\ref{fig:diagramencoder}).

\textbf{Word Embedding} maps words into a continuous space, which aligns words with similar intents, e.g. ``housing'' and ``dorm'', and separates words with different intents, e.g. ``Harvard'' and ``Cornell''.

The query term $t$ first goes through a standard embedding layer:
\begin{align*}
t \xrightarrow{emb} \vec{t}_{emb},
\end{align*}
which learns embeddings for all terms in the vocabulary.
The embeddings are fed into a highway network for better model capacity~\cite{highway}:
\begin{align}
\vec{t}_{hy} &= \text{highway}^t(\vec{t}_{emb}; g^{t}_{hy},  p^{t}_{hy}) \nonumber \\
&= g^{t}_{hy}(\vec{t}_{emb}) \times p^{t}_{hy}(\vec{t}_{emb}) + (1-g^{t}_{hy}(\vec{t}_{emb}))\times \vec{t}_{emb}. \label{eq.highway}
\end{align}
% $\vec{t}_{hy}$ is the highway embedding of term $t$.
% It includes a projection $p^t_{hy}$ and a gate $g^{t}_{hy}$:
It includes a non-linear projection:
\begin{align}
p^{t}_{hy} &= \text{relu}(w_{p^t_{hy}} \times \vec{t}_{emb}),
\end{align}
and a gate which controls the projection:
\begin{align}
g^{t}_{hy}(\vec{t}_{emb}) &= \text{sigmoid}(w_{g^t_{hy}} \times \vec{t}_{emb}).
\end{align}
Relu and sigmoid are activations; $w_{p^t_{hy}}$ and $w_{g^t_{hy}}$ are parameters.

%char representation
\textbf{Character-Aware Embedding.}
Search traffic follows a long tail distribution: many query terms only appear a few times or never before for reasons such as misspelling.
To help alleviate out-of-vocabulary issues, we employ a convolutional neural network over character embeddings to help represent rare words~\cite{huang2013learning, kim2016character}: the character-aware embeddings of ``retreval'' and ``retrieval'' can be similar because of shared characters, but the word embedding of ``retreval'' may not be well trained due to its low frequency caused by spelling errors.

The characters $C=\{c_1,...,c_j,...,c_m\}$ of term $t$ are embedded by an embedding layer, a Convolution layer, and a highway network.
\begin{align}
c_j & \xrightarrow{emb} \vec{c}_{j}, \label{eq.charemb}\\
\vec{t}_C &= \text{max-pool}(\texttt{CNN}(c_1,...,c_j,...,c_m)), \label{eq.charcnn} \\
\vec{t}_{C_{hy}} &= \text{highway}_2^C(\vec{t}_C; g_{hy1}^C, p_{hy1}^C, g_{hy2}^C, p_{hy2}^C). \label{eq.char-emb}
\end{align}
Eq.~\ref{eq.charemb} embeds each character;
Eq.~\ref{eq.charcnn} composes the embeddings of character n-grams and max-pools them to $\vec{t}_C$; 
Eq~\ref{eq.char-emb} is the same as Eq.~\ref{eq.highway} but has one more layer as the character vocabulary is smaller and can use more added capacity.

The word embedding  (Eq.~\ref{eq.highway}) and the character-aware embedding (Eq.~\ref{eq.char-emb}) are concatenated to the final term embedding:
\begin{align}
\vec{t} &= \vec{t}_{hy} \frown  \vec{t}_{C_{hy}}. \label{eq.mergechar}
\end{align}

\textbf{Mix Encoder} composes the word embeddings  $\{\vec{t}_1,...,\vec{t}_i,...,\vec{t}_n\}$ into the query encoding $\vec{\texttt{GEN}}(q)$.
It combines a sequential encoder to handle order sensitive queries, e.g. ``horse racing'' and ``racing horse'', and a bag-of-words model to model order insensitive queries, e.g. ``Cambridge MA'' and ``MA Cambridge''.

The sequential encoder is a one-layer bi-directional GRU:
\begin{align}
\{\overleftarrow{h_n}, \overrightarrow{h_n}\} &= \text{Bi-GRU}(\vec{t}_1,...,\vec{t}_i,...,\vec{t}_n). \label{eq.rnn}
\end{align}
$\overleftarrow{h_n}$ and $\overrightarrow{h_n}$ are the last states of the two directions.

\texttt{GEN} \texttt{Encoder} combines the Bi-GRU with the average of query word embeddings (bag-of-words) using a residual layer: 
% The Bi-GRU encoding are then composed with the bag-of-words model:
\begin{align}
\vec{\texttt{GEN}(q)} &= \text{tanh}(\vec{q}_{cat} + \text{relu}(w^q_{rs} \times \vec{q}_{cat} )), \label{eq.gin}\\
\vec{q}_{cat} &=  \overleftarrow{h_n} \frown \overrightarrow{h_n}\frown \frac{1}{n}\sum_{i}\vec{t}_i,  \label{eq.cat} 
\end{align}
where $w^q_{rs}$ is the parameter. $\vec{\texttt{GEN}(q)}$ is the encoding of $q$.

\subsection{Learning}
\label{sec:learning}
The distributed representations generated by \texttt{GEN} \texttt{Encoder} are determined by the information it learns from, which comes in two learning phases.
The first leverages user clicks as weak supervision to capture user intent; the second uses a combination of paraphrase labels in a multi-task setting to improve generality.

\input{table/flowers.tex}

\textbf{Weakly Supervised Learning.} 
Understanding user intents is difficult as they are not explicitly stated by the user. User clicks, however, have been widely used as implicit feedback of what a user wants~\cite{croft2010search}.
The first learning phase follows this intuition and assumes that queries with similar user clicks have similar user intents. It trains \texttt{GEN} \texttt{Encoder} to produce similar embeddings for such co-click queries and distinguishes those without shared clicks.

Let ($q$, $q^*$) be a pair of co-click queries sampled from the search log, 
the loss function of the first phase is
\begin{align}
l_{\text{co-click}} =& \sum_{q} \frac{1}{1 + \exp(\cos(\vec{\texttt{GEN}(q)}, \vec{\texttt{GEN}(q^*)}))} \label{eq.pos-loss}\\
&-\frac{1}{1 + \exp(\cos(\vec{\texttt{GEN}(q)}, \vec{\texttt{GEN}(q^-)}))}. \label{eq.neg-loss} 
\end{align}
This pairwise loss function trains the encoder to improve the cosine similarities between positive query pairs ($q, q^*$) and reduce those between negative pairs $(q, q^-)$.

Picking $q^-$ at random includes many trivial negatives that are easy to distinguish and less informative.
\texttt{GEN} \texttt{Encoder} uses noise-contrastive estimation (NCE) to pick adversarial negatives~\cite{mnih2013learning}:
\begin{align}
q^- &= \text{argmax}_{q' \in \text{batch}} \cos(\vec{\texttt{GEN}(q)}, \vec{\texttt{GEN}(q'))}.   \label{eq.nce}
\end{align}
It picks the negative $q^-$ in the current training batch that is the most similar to $q$ at the current learned version of \texttt{GEN} \texttt{Encoder}.

\textbf{Multi-Task Learning.} 
The weak supervision signals from co-clicks inevitably includes noise and may be biased towards the search engine's existing ranking system.
To learn a generic intent representation, the second learning phase adds two more human labeled paraphrase classification tasks, one on queries and one on questions, in a multi-task learning setting.

The query paraphrase dataset includes web search query pairs and expert labels: $\{q_i,q'_i, y_i\}$; $y_i=+1$ if $(q_i, q'_i)$ share the exact same intent (paraphrase) and $y_i=-1$ otherwise.
The question paraphrase dataset is the same except it includes natural language questions $\{qe_i,qe'_i, y_i\}$.

\texttt{GEN} \texttt{Encoder} is trained using logistic regression with cosine similarities as the loss:
\begin{align}
l_{\text{q-para}} &= \sum_i \frac{1}{1 + \exp(y_i \times \cos(\vec{\texttt{GEN}(q_i)}, \vec{\texttt{GEN}(q'_i)}))}, \\
l_{\text{qe-para}} &= \sum_j \frac{1}{1 + \exp(y_j \times \cos(\vec{\texttt{GEN}(qe_j)}, \vec{\texttt{GEN}(qe'_j)}))}. 
\end{align}

The two tasks are combined with co-click weak supervisions using multi-task learning:
\begin{align}
l_{\text{multi-task}} &= l_{\text{co-click}} + l_{\text{q-para}} + l_{\text{qe-para}}.
\end{align}

The first phase learns end-to-end from the large scale co-click signals from Bing search logs.
The second phase fine-tunes the model with high-quality but expensive human labels.

%% file: figure/GENDiagram.tex
\pdfoutput=1
\begin{figure*}[h]
\centering
 \begin{subfigure}{0.47\textwidth}
    \includegraphics[scale=0.29]{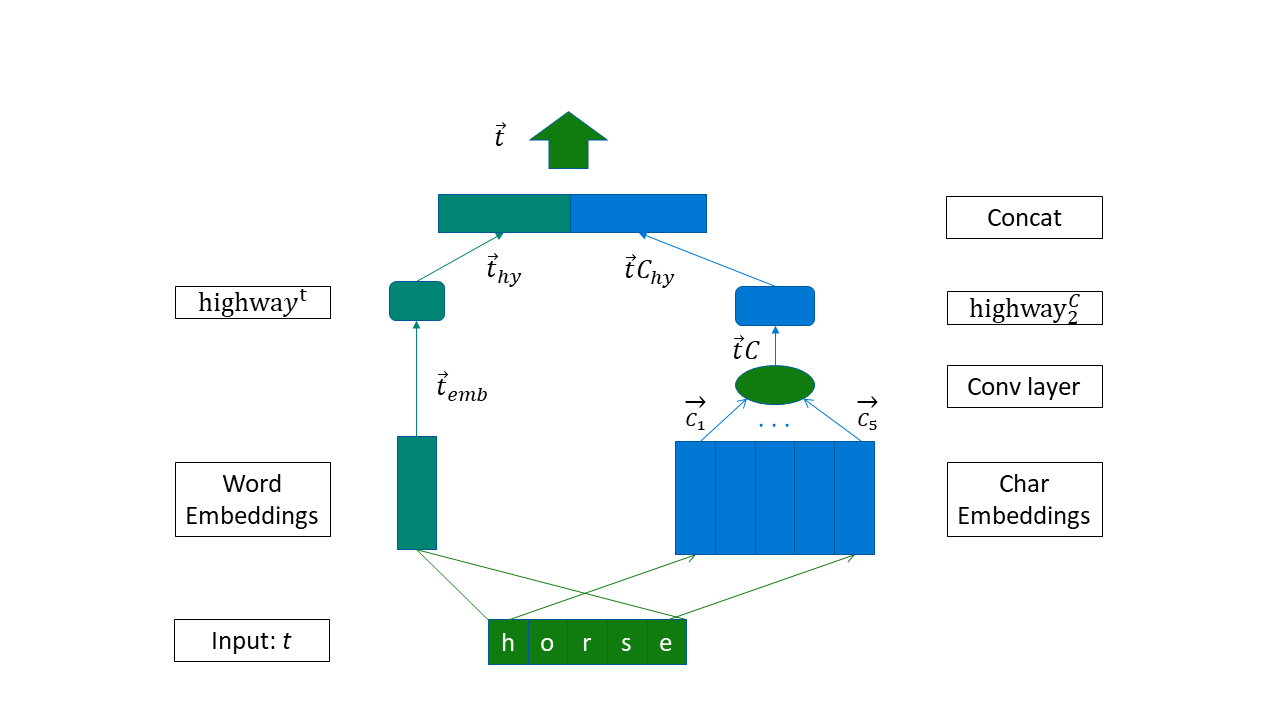}
    \caption{Word and Character Embedding\label{fig:diagramembedding}}
  \end{subfigure} \hfill
   \begin{subfigure}{0.47\textwidth}
    \includegraphics[scale=0.29]{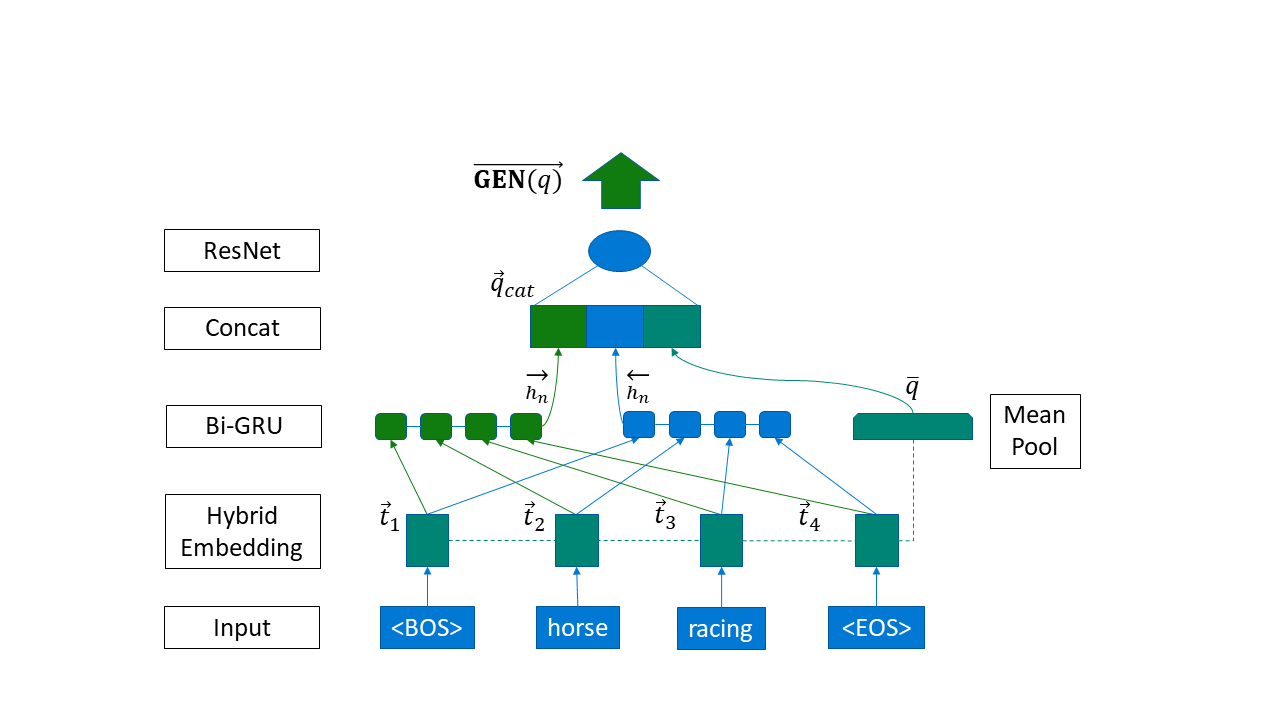}
    \caption{Mix Encoder\label{fig:diagramencoder}}
  \end{subfigure}\hfill
  \caption{GEN Encoder Architecture\label{fig:gendiagram}}
\end{figure*}

%% file: table/flowers.tex
\pdfoutput=1
\begin{table*}[th]
\centering
\caption{Statistics 
and Examples of the Three Intent Similarity Datasets. 
\label{tab:flowers} \label{tab:exg}}
\begin{tabular}{c|c|c|c|c} \hline \hline
\textbf{Data (\# Target Q)} & \textbf{Label} & \textbf{Count} & \multicolumn{2}{c}{\bf{Example Pair}} \\  \hline
\multirow{3}{*}{\texttt{General} (3,773)} &  Good & 2,720 & ``evaluation of instant checkmate'' & ``review of instant checkmate''  \\
& Fair & 2,609 & ``explorer how to export bookmarks''&  ``how to export bookmarks''  \\
 &    Bad & 9,804 & ``fantasy football contract leagues''& ``fantasy football league trophies''    \\ \hline
\multirow{3}{*}{\texttt{Easy} (2,864)}  &  Good  & 14,180 & ``cream of mushroom soup recipe''& ``cream mushroom soup recipe''  \\
& Fair &7,094 & ``louisiana cdl manual''& ``louisiana cdl training manual''  \\
 & Bad & 23,027 & ``how dose yelp work''& `` how long does yelp work'' \\ \hline
\multirow{2}{*}{\texttt{Hard} (385)}& Good & 193 & ``how to get rid of pimples''&    ``a fast way to get rid of pimples''  \\
& Bad  & 233 & ``horse racing''&  ``racing horse'' \\ \hline \hline
\end{tabular}
\end{table*}

%% file: Experiment.tex
\pdfoutput=1
\input{table/training.tex}
\input{table/detail}

\section{Intrinsic Evaluations}

Before demonstrating the potential of \texttt{Gen Encoder} in downstream applications, our first set of experiments directly evaluate the quality of the learned intent representations using \emph{intrinsic} evaluations.

\subsection{Experimental Methodologies}
\label{sec:exp}

This section first describes the experimental methodologies, including the task, training data, baselines, and implementation details.

\textbf{Evaluation Tasks.} 
Our experiments follow the common intrinsic evaluations on word embeddings~\cite{schnabel2015evaluation,finkelstein2002placing, wang2018glue} and evaluate the intent representations by their ability to \emph{model query similarities at the user intent level via their distances in the representation space}. 
Three \texttt{Intent} \texttt{Similarity} modeling datasets are used. 

\texttt{General} is the main dataset. 
It first samples a set of target queries from Bing search log. Then for each target query, several candidate queries appearing frequently with it in sessions are sampled. Human experts label the candidate queries based on their intent similarity with the target query in three scales: ``Good'', meaning they have the same intent and require the same search results; ``Fair'', which means the two queries are similar and some documents can satisfy both; ``Bad'' means the users are looking for different things for the two queries. Thus this is a graded scale of query intent similarity, i.e. a ranking based problem for evaluation.

\texttt{Easy} is an easier dataset. It focuses on trivial word level variations and spelling errors. It is also a ranking dataset.

\texttt{Hard} is the ``adversial'' dataset. It includes the failure cases from previously developed query representations in our system. It is a classification dataset with ``Good'' and ``Bad'' labels on query pairs.

Table~\ref{tab:flowers} lists the details and examples of the three datasets. They are sampled differently and there is little overlap between them.

\textbf{Evaluation Strategy.} 
For all evaluated representation models, they are first used to represent the queries. Then the distances between the query pairs' representations are evaluated against their similarity labels.
No information from \texttt{Intent} \texttt{Similarity} queries or labels is allowed to be used in model training or validation. 

The ranking datasets (\texttt{General} and \texttt{Easy}) are evaluated by NDCG with no cut-off.
The classification dataset (\texttt{Hard}) uses AUC score. The statistically significant differences between methods are tested using permutation (Fisher's randomization) test with $p<0.05$.

\input{table/overall}

\textbf{Training Data.} The two phase training of \texttt{GEN} \texttt{Encoder} uses co-click queries, query paraphrases, and question paraphrases.

\texttt{Co-click Queries} are sampled from six month of Bing search logs. 
Queries that lead to clicks on the same URL are collected to form co-click queries. 
URLs with more than five unique click queries are filtered out because they may reflect more than one intent.
The sample includes about 200 Million co-clicked query groups.

\texttt{Query Paraphrasing} contains about 0.8 million manually labeled query pairs with whether they are paraphrases of each other.

\texttt{Question Paraphrasing} has about 0.4 million question pairs manually labeled with whether they are paraphrases of each other.

Table~\ref{tab:training} lists the statistics of the three datasets.

\textbf{Baselines} include various representation methods, including bag-of-words, relevance-based embeddings,
representation-based neural ranking models, and other generic text representations.

Bag-of-word baselines include the following.
\begin{itemize}
\item \texttt{TF-IDF BOW} is the classic discrete bag-of-words based query representation. Its IDF is calculated from the search log.
\item \texttt{Emb BOW (d)} is the embedding-based language model~\cite{zamani2016estimating}. It uses GloVe embeddings pre-trained on documents~\cite{pennington2014glove}.
\item \texttt{Emb BOW (q)} is the embedding-based language model~\cite{zamani2016estimating}, with embeddings trained on search queries~\cite{nalisnick2016improving}.
\end{itemize}

Recent research suggests that distributed representations trained on pseudo relevance feedback are more effective in search tasks~\cite{zamani2017relevance,diaz2016query}. We use an enhanced version of relevance language model~\cite{zamani2017relevance} (\texttt{RLM+}), which we trained on relevance feedback (user clicks) which is better than pseudo relevance feedback, and then average the learned embeddings of query words to the query embeddings~\cite{zamani2016estimating}. It includes the following two versions.

\enlargethispage{\baselineskip}
\begin{itemize}
\item \texttt{RLM+ (t)} is trained on the query and clicked title pairs.
\item \texttt{RLM+ (u)} is trained on the query and clicked URL pairs.
\end{itemize}

Representation based neural ranking baselines include:
\begin{itemize}
\item \texttt{CDSSM}, a representation-based neural ranking model with character n-grams and CNN's~\cite{shen2014learning};
\item \texttt{Seq2Seq}, the Bi-GRU encoder-decoder model widely used in sequence to sequence learning.
\end{itemize}
\texttt{CDSSM} and \texttt{Seq2Seq} were both trained on query-clicked titles and query-clicked URL's, resulting in four neural ranking models.

Generic text representation baselines include the following:
\begin{itemize}
\item \texttt{BERT Encoder} is the encoder part of pre-trained BERT. The best performing setup in our testing is selected, which is the average of BERT base version's last layer~\cite{BERT}.
\item \texttt{USE}: the Universal Sentence Encoder trained on various language understanding tasks by multi-task learning~\cite{cer2018universal}.
\end{itemize}

The publicly released versions of \texttt{BERT}\footnote{https://github.com/google-research/bert} and \texttt{USE}\footnote{https://tfhub.dev/google/universal-sentence-encoder/2} are used as they are, in order to provide a better understanding of the effects of different training signals: surrounding text (\texttt{BERT}), language semantic understanding tasks (\texttt{USE}), and user clicks (\texttt{GEN} \texttt{Encoder}).
Fine-tuning or learning deeper architectures using our training data or combining these approaches are reserved for future research.

We implemented all other baselines following best practises suggested in previous research.
All methods, if applicable, were trained using the same search log sample, the same scale of training data, and the same hyper-parameter settings. They are then evaluated exactly the same with \texttt{GEN} \texttt{Encoder}.

\texttt{RLM+} is the most related query representation method designed for information retrieval tasks~\cite{zamani2017relevance}. It performed better than the locally trained word embeddings~\cite{diaz2016query} and discrete PRF expansions~\cite{lavrenko2001relevance}, making it the main IR baseline in our evaluations.
\texttt{USE} is the most related generic representation baseline and has shown strong performances on language understanding tasks.

\textbf{Implementation Details} The parameters of \texttt{GEN} \texttt{Encoder} and their dimensions are listed in Table~\ref{tab:para}. 
Training uses Adam optimizer, 1e-4 learning rate, and 256 batch size.
All parameters are learned end-to-end. Learning in each phase is concluded base on validation loss.
On a typical GPU, the first phase takes about 300 hours per epoch and converges after 1 epoch; the second phase takes one hour per epoch and several epochs to converge. The multi-task learning phase randomly mixes the training data from three datasets in each mini-batch.
On a typical GPU it takes about 15 milliseconds to infer the \texttt{GEN} \texttt{Encoding} for an average query.

%% file: table/training.tex
\pdfoutput=1
\begin{table*}[th]
\centering
\caption{Datasets used to train \texttt{GEN} \texttt{Encoder} in the two-stage setting: first weak-supervision and then multi-task learning.
\label{tab:training}}
\begin{tabular}{c|c|c|c|c|c}
\hline \hline
\textbf{Stage} & \textbf{Dataset} & \textbf{Training} & \textbf{Validation}  & \textbf{Testing} & \textbf{Labels} \\ \hline 
Weak-Supervision & Co-click Queries & $\approx$200M groups & $\approx$10K groups & -- & 50\%+/50\%- \\ \hline
\multirow{3}{*}{Multi-Task}  & Co-click Queries & $\approx$400K groups & $\approx$10K groups & -- & 50\%+/50\%- \\
& Query Paraphrases & $\approx$800K pairs &$\approx$10K pairs & $\approx$10K pairs& $\approx$25\%+/75\%-  \\
& Question Paraphrases & $\approx$350K pairs & $\approx$10K pairs & $\approx$10K pairs &$\approx$30\%+/70\%- \\ \hline \hline
\end{tabular}
\end{table*}

%% file: table/detail.tex
\pdfoutput=1
\begin{table}[t]
\centering
    \caption{\texttt{GEN} \texttt{Encoder} Parameters. \label{tab:para}}
    \begin{tabular}{c|c|l} \hline \hline
    \textbf{Parameter} & \textbf{Dimension} & \textbf{Description} \\ \hline
    $\vec{t}_{emb}$ & 1M*200 & Word Embedding \\
    $w_{g^t_{hy}}$ & 200*200 & Word Highway Gate \\
    $w_{p^t_{hy}}$ & 200*200 & Word highway Projection \\ \hline
    $\vec{c}$ & 1000*200 & Character Embedding \\
    CNN & [5,100] & Char-Ngram CNN \\ 
     $w_{g^C_{hy1}}$,  $w_{g^C_{hy2}}$ & 100*100 & Char Highway Gate \\
     $w_{p^C_{hy1}}$,  $w_{p^C_{hy2}}$ & 100*100 & Char Highway Projection \\ \hline \hline
     Bi-GRU Layer & 512 & RNN Encoder  \\ 
     $w_{rs}^q$ & 1324*100 & ResNet on Bi-GRU\&BOW \\ \hline \hline
    \end{tabular}
    
\end{table}

%% file: table/overall.tex
\pdfoutput=1
\begin{table*}
\centering
\caption{Performances on Query Intent Similarity Modeling. \label{tab:overall}
% The evaluation metrics for each dataset are in brackets. 
Relative difference over \texttt{TF-IDF} \texttt{BOW} are presented in percentages. 
$\dagger, \ddagger, \mathsection, \mathparagraph$ marks the statistical significant improvements over \texttt{TF-IDF BOW}$^\dagger$, \texttt{Emb} \texttt{BOW} (q)$^\ddagger$, \texttt{RLM+} \texttt{(title)}$^\mathsection$ and \texttt{USE}$^{\mathparagraph}$ (p<0.05).
\label{tab:eval_full}}
\begin{tabular}{l|lc|lc|lc|lc|lc}
\hline \hline
& \multicolumn{6}{c|}{\bf{Query Intent Similarity}} & \multicolumn{4}{c}{\bf{Paraphrase Classification}} \\ \hline
\bf{Method}& \multicolumn{2}{c|}{\bf{General (NDCG)}} & \multicolumn{2}{c|}{\bf{Easy (NDCG)}}& \multicolumn{2}{c|}{\bf{Hard (AUC)}}
 & \multicolumn{2}{c|}{\bf{Question (AUC)}}& \multicolumn{2}{c}{\bf{Query (AUC)}}
\\ \hline
\texttt{TF-IDF BOW} 
& $0.4969$ & --  & $0.8047$ & --  & $0.4740$ & --  & $0.6869$ & --  & $0.8992$ & -- \\ \hline
\texttt{Emb BOW (d)} 
 & ${0.4842}$ & $ -2.54\%  $ 

 & ${0.8567}^{\dagger }$ & $ +6.47\%  $ 

 & ${0.5059}$ & $ +6.73\%  $ 

 & ${0.7093}^{\dagger }$ & $ +3.26\%  $ 

 & ${0.9146}$ & $ +1.71\%  $ 

 \\
 \texttt{Emb BOW (q)}
 & ${0.4834}$ & $ -2.71\%  $ 

 & ${0.8583}^{\dagger \mathsection }$ & $ +6.66\%  $ 

 & ${0.5055}$ & $ +6.65\%  $ 

 & ${0.7111}^{\dagger }$ & $ +3.52\%  $ 

 & ${0.9186}$ & $ +2.15\%  $ 

\\\hline

\texttt{CDSSM (t)} 
 & ${0.3830}$ & $ -22.92\%  $ 

 & ${0.8237}^{\dagger }$ & $ +2.36\%  $ 

 & ${0.4476}$ & $ -5.58\%  $ 

 & ${0.5382}$ & $ -21.64\%  $ 

 & ${0.5973}$ & $ -33.57\%  $ 
 \\
\texttt{CDSSM (u)} 
 & ${0.3857}$ & $ -22.37\%  $ 

 & ${0.8281}^{\dagger }$ & $ +2.91\%  $ 

 & ${0.4700}$ & $ -0.84\%  $ 

 & ${0.5067}$ & $ -26.23\%  $ 

 & ${0.6112}$ & $ -32.03\%  $ 
 \\
\texttt{Seq2Seq (t)} 
 & ${0.4606}$ & $ -7.30\%  $ 

 & ${0.8593}^{\dagger \ddagger \mathsection }$ & $ +6.79\%  $ 

 & ${0.5300}$ & $ +11.81\%  $ 

 & ${0.7184}^{\dagger }$ & $ +4.58\%  $ 

 & ${0.8724}$ & $ -2.98\%  $ 
\\ 
\texttt{Seq2Seq (u)} 
 & ${0.4499}$ & $ -9.45\%  $ 

 & ${0.8584}^{\dagger \ddagger \mathsection }$ & $ +6.68\%  $ 

 & ${0.5106}$ & $ +7.71\%  $ 

 & ${0.7044}$ & $ +2.55\%  $ 

 & ${0.8721}$ & $ -3.02\%  $ 
 \\ 
 
 \hline

\texttt{RLM+ (t)} 
 & ${0.4985}^{\dagger \ddagger \mathparagraph }$ & $ +0.33\%  $ 

 & ${0.8570}^{\dagger }$ & $ +6.50\%  $ 

 & ${0.5036}$ & $ +6.24\%  $ 

 & ${0.7343}^{\dagger \ddagger }$ & $ +6.91\%  $ 

 & ${0.9562}^{\dagger \ddagger \mathparagraph }$ & $ +6.33\%  $ 
\\
\texttt{RLM+ (u)} 
 & ${0.4890}^{\ddagger }$ & $ -1.59\%  $ 

 & ${0.8592}^{\dagger \ddagger \mathsection }$ & $ +6.77\%  $ 

 & ${0.4938}$ & $ +4.17\%  $ 

 & ${0.7138}^{\dagger }$ & $ +3.91\%  $ 

 & ${0.9331}^{\dagger \ddagger }$ & $ +3.77\%  $ 
\\

 \hline

\texttt{BERT Encoder}
 & ${0.4643}$ & $ -6.56\%  $ 

 & ${0.8585}^{\dagger \ddagger \mathsection }$ & $ +6.69\%  $ 

 & ${0.4977}$ & $ +5.00\%  $ 

 & ${0.7352}^{\dagger \ddagger }$ & $ +7.04\%  $ 

 & ${0.9027}$ & $ +0.39\%  $ 

 \\
 
\texttt{USE}
 & ${0.4958}^{\ddagger }$ & $ -0.21\%  $ 

 & ${0.8635}^{\dagger \ddagger \mathsection }$ & $ +7.31\%  $ 

 & ${0.5675}^{\dagger \mathsection }$ & $ +19.72\%  $ 

 & ${0.7974}^{\dagger \ddagger \mathsection }$ & $ +16.09\%  $ 

 & ${0.9477}^{\dagger \ddagger }$ & $ +5.39\%  $ 
 
\\\hline
\texttt{GEN} \texttt{Encoder}
 & ${0.5244}^{\dagger \ddagger \mathparagraph }$ & $ +5.53\%  $ 

 & ${0.8688}^{\dagger \ddagger \mathsection \mathparagraph }$ & $ +7.97\%  $ 

 & ${0.6667}^{\dagger \ddagger \mathsection \mathparagraph }$ & $ +40.64\%  $ 

 & ${0.8486}^{\dagger \ddagger \mathsection \mathparagraph }$ & $ +23.55\%  $ 

 & ${0.9863}^{\dagger \ddagger \mathsection \mathparagraph }$ & $ +9.68\%  $ 

\\
 \hline \hline
\end{tabular}
\end{table*}

%% file: Evaluation.tex
\pdfoutput=1

\subsection{Evaluation Results}
\label{sec:eva}
Two experiments are conducted to study \texttt{GEN} \texttt{Encoder}'s accuracy and source of effectiveness using the intent similarity datasets.

\subsubsection{Accuracy on Modeling Intent Similarity}
 
% The performances of all compared methods are in 
Table~\ref{tab:eval_full} shows the overall evaluation results.
Baselines are grouped as bag-of-words (\texttt{BOW}), neural matching models (\texttt{CDSSM}, \texttt{Seq2Seq}), relevance-based embeddings (RLM+), and generic representations (\texttt{BERT}, \texttt{USE}).
For the generic representations,  performance on the Paraphrase Classification tasks are presented, but more for reference, as they were not fine-tuned on paraphrase labels.
All methods were evaluated on Query Intent Similarity exactly the same.

\input{table/ablation}

The embedding methods \texttt{Emb BOW} and the representation-based neural ranking methods (\texttt{CDSSM} and \texttt{Seq2Seq}) do not outperform \texttt{TF-IDF BOW}. Embeddings trained on surrounding texts or representations learned to match documents are not designed to represent search intents.
\texttt{RLM+} performs better than \texttt{BOW} models in the majority of datasets. 
Embeddings trained using relevance signals are more effective than embeddings trained on surrounding contexts in modeling user intent:
\texttt{RLM+} uses the same average of word embeddings with \texttt{Emb BOW}, but its embeddings are trained on the relevance feedback signals which better reflects information needs~\cite{zamani2017relevance}.

\texttt{USE} shows strong performances across the datasets, while \texttt{BERT} \texttt{Encoder}'s effectiveness is more mixed.
The two differ in two aspects.
First, \texttt{USE} is designed to learn general representations and focuses on building a universal representation space~\cite{cer2018universal}. In comparison, \texttt{BERT} is a sequence to sequence model; much of the model capacity resides in the cross-sequence multi-head attentions, instead of general representation~\cite{BERT}.
Second, \texttt{USE} leverages multi-task learning to improve the learned representation's ``Universal'' generalization ability~\cite{cer2018universal} while \texttt{BERT} focuses on single task.

\texttt{GEN} \texttt{Encoder} outperforms all baselines on all datasets.
Its advantages are significant, stable, and stronger on harder datasets.
It is the only one that significantly outperforms \texttt{TF-IDF (BOW)} on \texttt{General}, and outperforms all other methods on \texttt{Hard} with large margins (at least 17\%).
\texttt{GEN} \texttt{Encoder} is more effective on all intent similarity datasets compared to \texttt{RLM+}, the most related IR-style query representation method.
It also significantly outperforms \texttt{USE}, the previous state-of-the-art in general representation learning. 
The next experiment studies the reasons for these improvements.

\subsubsection{Source of Effectiveness}
This experiment studies the contribution of \texttt{GEN} \texttt{Encoder}'s training strategies and neural architecture.

\textbf{Learning Ablations.}
The top half of Table~\ref{tab:ablation} shows the model performances with different subsets of training data.
It varies the percentage of co-click data used, as shown on the left of the Method column, and the set of paraphrase labels used in the second phase, as shown on the right of the Method column.

The co-click weak supervision is the most important for \texttt{GEN} \texttt{Encoder}.
When no co-click is used (0\%), there is not sufficient data to learn a meaningful intent representation space.
When all co-click signals are used (100\%), \texttt{GEN} \texttt{Encoder} reaches its peak performances on \texttt{General} and \texttt{Easy} datasets, even without the second learning phase. The noise-contrastive estimation (NCE) also improves the accuracy with more informative negative training samples.

Multi-task learning is crucial to the model generalization and performances on \texttt{Hard}.
It improves the AUC on \texttt{Hard} by about $10\%$, \texttt{full} \texttt{model} vs.\ \texttt{(100\%, none)}.
On the other hand, if only one task is used in the second phase, \texttt{GEN} \texttt{Encoder} performs well on the corresponding task: (100\%, query) on Query Paraphrase and (100\%, question) on Question Paraphrase, but worse on all other tasks than only using the first phase \texttt{(100\%, none)}.

\textbf{Encoder Ablations.} The lower half of Table~\ref{tab:ablation} shows the performances of different encoder architectures: \texttt{Avg-Emb} is the average of word embeddings; \texttt{Bi-GRU} is the vanilla Bi-GRU on word embeddings; \texttt{no} \texttt{char-embedding} discards the character embedding component; \texttt{no highway} discards all highway layers.
All encoder components contribute; discarding any of them often reduces performances, mostly noticeably on \texttt{Hard} and \texttt{Question} \texttt{Paraphrase}.

This experiment demonstrates the importance of learning strategy in building generic intent representations.
Trained on the right data, a simple embedding (\texttt{Avg-Emb}) outperforms more complex models trained for other purposes.
\texttt{Avg-Emb} even reaches the performance of \texttt{USE}, which uses a deep transformer architecture and multi-task learning, but is not trained for search tasks. 

An interesting finding in our ablation study is that the encoder may not have sufficient capacity to consume weak supervision signals at this scale. Its accuracy plateaus with about 10\% co-click data (20 million).
The ability to learn from user clicks enables large scale weak supervision; to fully leverage the potential of this available scale may require training deep architectures more efficiently, which is a future research direction.

%% file: table/ablation.tex
\pdfoutput=1
\begin{table*}
\centering
\caption{Performances of \texttt{GEN} \texttt{Encoder} variations. 
Relative performances and statistically significant differences$^*$ are compared to the full \texttt{GEN} \texttt{Encoder}.
The varied training data in the two learning phases are listed in the brackets: (first, second).
\label{tab:ablation}
}
\begin{tabular}{l|lc|lc|lc|lc|lc}
\hline \hline
\multicolumn{11}{c}{\bf{Different Learning Strategies, Same Gen Encoder}} \\ \hline
& \multicolumn{6}{c|}{\bf{Query Intent Similarity}} & \multicolumn{4}{c}{\bf{Paraphrase Classification}} \\ \hline
\bf{Method}& \multicolumn{2}{c|}{\bf{Normal (NDCG)}} & \multicolumn{2}{c|}{\bf{Easy (NDCG)}}& \multicolumn{2}{c|}{\bf{Hard (AUC)}}
 & \multicolumn{2}{c|}{\bf{Question (AUC)}}& \multicolumn{2}{c}{\bf{Query (AUC)}}
\\ \hline

 \texttt{(1\%, none)} 
 & ${0.5136}^{* }$ & $ -2.05\%  $ 

 & ${0.8711}^{* }$ & $ +0.26\%  $ 

 & ${0.5781}^{* }$ & $ -13.28\%  $ 

 & ${0.7646}^{* }$ & $ -9.90\%  $ 

 & ${0.9273}^{* }$ & $ -5.98\%  $
 \\ 
  \texttt{(10\%, none)} 
 & ${0.5230}^{* }$ & $ -0.27\%  $ 

 & ${0.8719}^{* }$ & $ +0.36\%  $ 

 & ${0.6018}^{* }$ & $ -9.73\%  $ 

 & ${0.7832}^{* }$ & $ -7.71\%  $ 

 & ${0.9503}^{* }$ & $ -3.65\%  $
 \\

 \texttt{(100\%, none)}
 & ${0.5278}^{* }$ & $ +0.65\%  $ 

 & ${0.8734}^{* }$ & $ +0.52\%  $ 

 & ${0.6059}^{* }$ & $ -9.12\%  $ 

 & ${0.7795}^{* }$ & $ -8.15\%  $ 

 & ${0.9519}^{* }$ & $ -3.48\%  $ 
 \\ 
  \texttt{(100\% no NCE, none)}
 & ${0.5225}^{* }$ & $ -0.36\%  $ 

 & ${0.8711}^{* }$ & $ +0.26\%  $ 

 & ${0.5880}^{* }$ & $ -11.79\%  $ 

 & ${0.7694}^{* }$ & $ -9.34\%  $ 

 & ${0.9515}^{* }$ & $ -3.52\%  $

 \\\hline
\texttt{(0\%, query)}  
 & ${0.4273}^{* }$ & $ -18.51\%  $ 

 & ${0.8376}^{* }$ & $ -3.59\%  $ 

 & ${0.4943}^{* }$ & $ -25.85\%  $ 

 & ${0.5680}^{* }$ & $ -33.07\%  $ 

 & ${0.9532}^{* }$ & $ -3.35\%  $ 
 \\
\texttt{(0\%, question)}  
 & ${0.4232}^{* }$ & $ -19.30\%  $ 

 & ${0.8457}^{* }$ & $ -2.66\%  $ 

 & ${0.4373}^{* }$ & $ -34.40\%  $ 

 & ${0.6465}^{* }$ & $ -23.82\%  $ 

 & ${0.7612}^{* }$ & $ -22.81\%  $
 \\ 
 \texttt{(0\%, multi-task)}  
 & ${0.5101}^{* }$ & $ -2.72\%  $ 

 & ${0.8661}^{* }$ & $ -0.31\%  $ 

 & ${0.5220}^{* }$ & $ -21.70\%  $ 

 & ${0.7912}^{* }$ & $ -6.77\%  $ 

 & ${0.9725}^{* }$ & $ -1.40\%  $ 
 \\
 \hline

\texttt{(100\%, query)}  
 & ${0.5056}^{* }$ & $ -3.58\%  $ 

 & ${0.8667}^{* }$ & $ -0.24\%  $ 

 & ${0.5955}^{* }$ & $ -10.68\%  $ 

 & ${0.6969}^{* }$ & $ -17.88\%  $ 

 & ${0.9869}$ & $ +0.06\%  $
 \\
\texttt{(100\%, question)} 
 & ${0.5164}^{* }$ & $ -1.52\%  $ 

 & ${0.8679}^{* }$ & $ -0.10\%  $ 

 & ${0.5917}^{* }$ & $ -11.24\%  $ 

 & ${0.8496}$ & $ +0.11\%  $ 

 & ${0.9585}^{* }$ & $ -2.81\%  $
\\
\hline
\texttt{full model} 
 & $0.5244$ & --  & $0.8688$ & --  & $0.6667$ & --  & $0.8486$ & --  & $0.9863$ & --
\\ 

\hline \hline
\multicolumn{11}{c}{\bf{Different Encoder Architecture, Same Two-Phase Learning}}  \\
\hline
& \multicolumn{6}{c|}{\bf{Query Intent Similarity}} & \multicolumn{4}{c}{\bf{Paraphrase Classification}} \\ \hline
\bf{Method}& \multicolumn{2}{c|}{\bf{Normal (NDCG)}} & \multicolumn{2}{c|}{\bf{Easy (NDCG)}}& \multicolumn{2}{c|}{\bf{Hard (AUC)}}
 & \multicolumn{2}{c|}{\bf{Question (AUC)}}& \multicolumn{2}{c}{\bf{Query (AUC)}}
\\ \hline

\texttt{Avg-Emb} 
 & ${0.5081}^{* }$ & $ -3.11\%  $ 

 & ${0.8612}^{* }$ & $ -0.87\%  $ 

 & ${0.5670}^{* }$ & $ -14.95\%  $ 

 & ${0.8286}^{* }$ & $ -2.37\%  $ 

 & ${0.9620}^{* }$ & $ -2.46\%  $
\\
\texttt{Bi-GRU} 
 & ${0.5055}^{* }$ & $ -3.59\%  $ 

 & ${0.8633}^{* }$ & $ -0.63\%  $ 

 & ${0.5494}^{* }$ & $ -17.59\%  $ 

 & ${0.7665}^{* }$ & $ -9.68\%  $ 

 & ${0.9681}^{* }$ & $ -1.84\%  $
\\

\hline
\texttt{no char-embedding}
 & ${0.5182}^{* }$ & $ -1.17\%  $ 

 & ${0.8726}^{* }$ & $ +0.43\%  $ 

 & ${0.6292}$ & $ -5.61\%  $ 

 & ${0.7797}^{* }$ & $ -8.13\%  $ 

 & ${0.9546}^{* }$ & $ -3.21\%  $ 
 \\ 
\texttt{no highway} 
 & ${0.5254}^{* }$ & $ +0.20\%  $ 

 & ${0.8686}^{* }$ & $ -0.02\%  $ 

 & ${0.6172}$ & $ -7.43\%  $ 

 & ${0.7459}^{* }$ & $ -12.11\%  $ 

 & ${0.9495}^{* }$ & $ -3.72\%  $
 \\ 
 
 \hline
 
 \texttt{full model}
 & $0.5244$ & --  & $0.8688$ & --  & $0.6667$ & --  & $0.8486$ & --  & $0.9863$ & --

\\ \hline

\hline \hline
\end{tabular}
\end{table*}

%% file: Applications.tex
\pdfoutput=1

\section{Alleviating the Tail Sparsity}
\label{sec:ann}
% \section{Soft Retrieval in GEN Space}
Much of a search engine's intelligence comes from its users.
User click behavior from previous observations of a query is one of the most effective relevance signals~\cite{agichtein2006improving}; previous clicked queries on a document is among the most informative document representation in relevance ranking~\cite{Zamani2018NeuralRM}.
However, the long tail nature of search traffic means many queries are rarely, if ever, observed, 

This section demonstrates how \texttt{GEN} \texttt{Encoder} alleviates the long tail challenge by mapping a rare query to previously observed queries sharing the same intent by using approximate nearest neighbor (ANN) search. 
We first describe the setup of our experiment, followed by quantitative and qualitative analyses of the tail alleviation, and then discuss how this phenomenon improves the coverage of our online question answering capabilities.

\input{table/anntable}

\textbf{ANN Index and Queries.} We built an ANN index of the \texttt{GEN} encodings of 700 million queries sampled from six months, using the HNSW algorithm~\cite{malkov2018efficient}.
In a typical parallel computing environment (feasible in academic settings), the index performs an ANN lookup within at most 10 milliseconds.
For this study, we randomly sampled one million queries occurring on one day some time after the ANN index is built, with navigational and adult queries filtered out. About 5\% of them are head (appear more than $2^{15}$ times), 40\% are torso ($2^5$ to $2^{15}$), and 55\% are tail queries ($\leq 2^4$).
We retrieved the top ten nearest neighbors for each query within varying cosine distance radius (0.15, 0.10, or 0.05).

\textbf{ANN Coverage and Accuracy.}
Table~\ref{tab:ann} shows the statistics of the ANN search results for the one million queries.
The Co-Intent $\%$ evaluates the fraction of ANN queries sharing the exact same intent as the search query. We asked human judges to evaluate the retrieved nearest neighbors of 100 randomly sampled queries, comprised of 3 head, 47 torso, and 50 tail queries.
Three experts labeled them with an average of 0.717 Cohen's Kappas (high agreement). Disagreements were resolved by majority.

The ANN search covers most  head and torso queries. 
Table Table~\ref{tab:ann} shows the coverage on tail queries is significantly more sparse, as expected, but still half of them have ANN queries within radius of 0.15, with average $2-3$ ANN per query. However, not all of the nearest neighbors are co-intent queries (about 47\% are in the tail), so each of the $57.9\%$ of tail queries which have neighbors will have about 1.37 ($2.91\times47\%$) co-intent neighbors.
This result confirms that many tail queries are rare ways of expressing a search intent that has been expressed by another query before~\cite{downey2007heads}, and that ANN search in the \texttt{GEN} encoder space (\texttt{ANN-GEN}) can efficiently retrieve such co-intent queries for a large fraction of tail queries.

\textbf{Alleviating the Tail.} 
A direct impact of \texttt{ANN-GEN} is to alleviate the sparsity of tail queries by combining observations from the semantically similar queries found using ANN search. For example, the user click behaviors from approximate neighbors of a tail query can be used to improve its relevance ranking. 

\input{figure/smoothtail.tex}

Figure~\ref{fig:smooth} illustrates how bringing in approximate neighbors of a search query can make the tail less sparse. The Query Frequency counts the occurrences of tail queries in a very large sample from the span of six months right before the day the query was sampled from. The ANN Frequency is the total number of occurrences if the query's approximate neighbors are introduced and their observations are added to the search query's frequency. Since we filtered out navigational traffic, the line with square marks in Figure~\ref{fig:smooth} shows that about $39\%$ of our total query sample never appeared in the previous six month window, and thus there is no feedback signals for them at all.  ANN alleviated the sparsity on extreme rare queries significantly: the $\approx$40\% of queries which are unseen in the log drops by half at the 0.15 radius threshold (Figure~\ref{fig:smoothann}), or by $\approx$35\% if we account for neighbors which are non-co-intent (Figure~\ref{fig:smoothacc}).

\input{figure/density.tex}

\input{table/pairclasses}
\input{table/caseStudy.tex}

\textbf{Influences in Online Production.}
This ANN search with \texttt{GEN} \texttt{Encoder} has been deployed in various components of Bing search systems.
One of its many applications is to improve the coverage of the search engine's online question answering by aligning unanswerable questions to their answerable semantically equivalent \texttt{ANN-GEN} questions. The coverage of an important fraction of online QA pipeline was {\em doubled} without loss of answer quality.

\section{Understanding Search Behavior}
\label{sec:dist}
Another advantage of distributed representations is that the distance in the embedding space may reflect certain meaningful relationships between the points. This section analyzes the distances in \texttt{Gen} \texttt{Encoder}'s representation space, and then shows the distances between query pairs reflect interesting behaviors in user sessions.

\textbf{Data.}
The dataset used is one million search sessions sampled from one day's search logs. Sessions are delineated by a 30-minute gap.
Standard non-navigational and work-safe filters are applied. Sessions with less than three queries or no click are filtered.

\textbf{Distances Distributions.}
Figure~\ref{fig:distance} plots the distributions of distances of query pairs in \texttt{Emb BOW (q)}, \texttt{BERT Encoder}, and \texttt{GEN} \texttt{Encoder}.
As expected, query pairs appearing together in sessions are more similar.
However, the distributions of distances in these three methods differ considerably: The adjacent query pairs in sessions (query reformulations) follow a strong and flat bi-modal distribution in \texttt{GEN} \texttt{Encoder}, very different from the  distributions from previous embedding methods.

\textbf{Human Analyses.}
To understand this bi-modal distribution, we conduct human analyses on the relations between these adjacent queries.
One hundred adjacent query pairs are randomly sampled from these sessions. Three judges label them into four categorizes of information seeking behaviors, following previous research in session understanding~\cite{hassan2014struggling, sessiontrack}
\begin{enumerate}
\item \texttt{Topic Change}: The two queries are unrelated.
\item \texttt{Explore}: The second query explores within topic space of the previous query.
\item \texttt{Specify}: The second query narrows down the intent of the previous query.
\item \texttt{Paraphrase}: The second query has the exact same intent of the previous query.
\end{enumerate}
They agree at 0.64 Cohen's Kappa, which is a reasonable high agreement when labeling four classes.
Disagreements were resolved by majority. 
Ties are broke by a fourth judge.

Table~\ref{tab:catcor} lists the correlations between representation similarities and human labels. 
Table~\ref{tab:casestudy} lists some example query pairs.
\texttt{GEN} \texttt{Encoder} correlates the most with human labels.
As shown in the Related and Middle columns,
Though all methods seem able to differentiate ``Topic Change'' from the rest,
\texttt{GEN} \texttt{Encoder} does the best at identifying ''Explore'' and ''Specify'' categories,  which  capture interesting information seeking behaviors in search sessions.

This result further demonstrates the quality of the \texttt{GEN} \texttt{Encoder}'s generic intent representation space
and suggests its future applications in analyzing and understanding user's information seeking behavior.

%% file: table/anntable.tex
\pdfoutput=1
\begin{table}
\centering
\caption{Statistics of ANN at different distance radius (0.15, 0.10, 0.05): percent of queries that have ANN neighbors (Coverage), average number of ANN neighbors (\# Neighbor) among covered queries, and the fraction of ANN neighbors sharing same user intents (Co-Intent \%).
\label{tab:ann}}

\begin{tabular}{c|p{0.42cm}p{0.42cm}p{0.42cm}|p{0.42cm}p{0.42cm}p{0.42cm}|p{0.42cm}p{0.42cm}p{0.42cm}} \hline \hline
&\multicolumn{3}{c|}{\textbf{Coverage \%}} 
&\multicolumn{3}{c|}{\textbf{\# Neighbor}} 
&\multicolumn{3}{c}{\textbf{Co-Intent \%}} \\ \hline
& \textbf{0.15}
& \textbf{0.10}
& \textbf{0.05} 
& \textbf{0.15}
& \textbf{0.10}
& \textbf{0.05} 
& \textbf{0.15}
& \textbf{0.10}
& \textbf{0.05}  \\ \hline

Head & 96.1 & 90.0 & 79.8 & 5.05 & 4.92 & 4.32 & 92 & 100 & 100 \\
Torso & 88.3 & 75.2 & 59.7 & 3.77 & 3.57 & 3.17 & 80 & 90 & 99 \\
Tail & 57.9 & 32.9 & 15.9 & 2.91 & 2.74 & 2.32 & 47 & 59 & 80\\ \hline \hline
\end{tabular}

\end{table}

%% file: figure/smoothtail.tex
\pdfoutput=1
\begin{figure}[t]
 \begin{subfigure}{0.23\textwidth}
\includegraphics[scale=0.27]{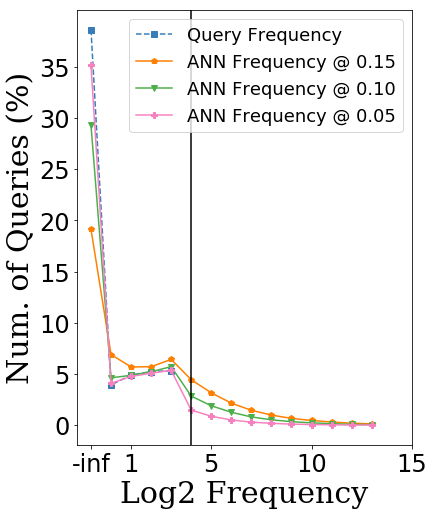}
    \caption{All Frequency\label{fig:smoothann}}
    \label{fig:demo1}
  \end{subfigure} \hfill
  \begin{subfigure}{0.23\textwidth}
    \includegraphics[scale=0.27]{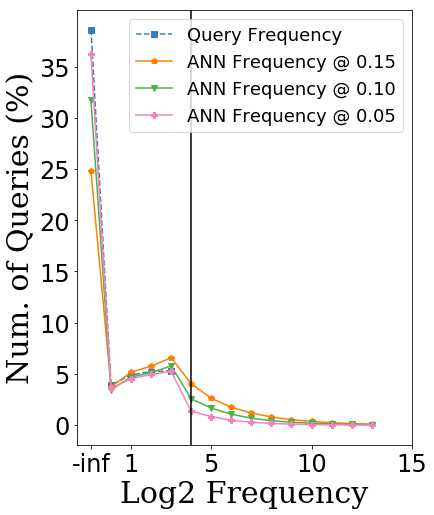}
    \caption{Penalized Frequency\label{fig:smoothacc}}
    \label{fig:demo2}
  \end{subfigure}\hfill
  \caption{The  distributions of tail query frequency and their ANN frequency at different neighbor radius.
  Queries and ANNs are binned by their frequencies in the six-month period before their date.  X-axis marks the log$_2$ frequency; -inf indicates unseen queries. ANN Frequency in (b) is penalized by co-intent accuracy (e.g. 0.47) from Table $\ref{tab:ann}$.
  Y-axis is fraction of corresponding bin in total traffic.
  The tail range ($\leq2^4$) are marked by vertical lines.
  \label{fig:smooth}
  }
\end{figure}

%% file: figure/density.tex
\pdfoutput=1
\begin{figure*}[t]
    \centering
    \begin{subfigure}[h]{0.33\textwidth} \centering
        \includegraphics[scale=0.25]{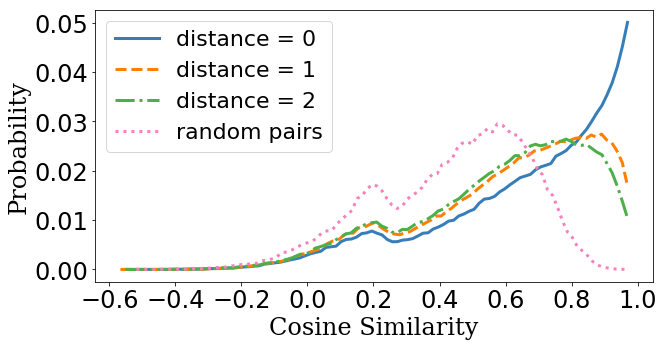}
        \caption{Emb BOW (q)} \hfill
        % \label{fig:my_label}
    \end{subfigure} \hfill
     \begin{subfigure}[h]{0.33\textwidth} \centering
        \includegraphics[scale=0.25]{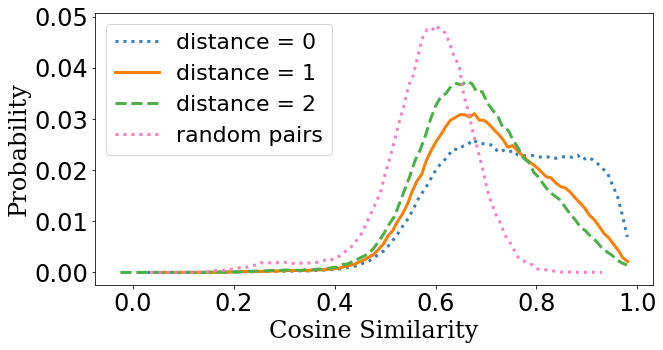}
        \caption{BERT Encoder}  \hfill
        % \label{fig:densityPlot}
    \end{subfigure} \hfill
    \begin{subfigure}[h]{0.33\textwidth} \centering
        \includegraphics[scale=0.25]{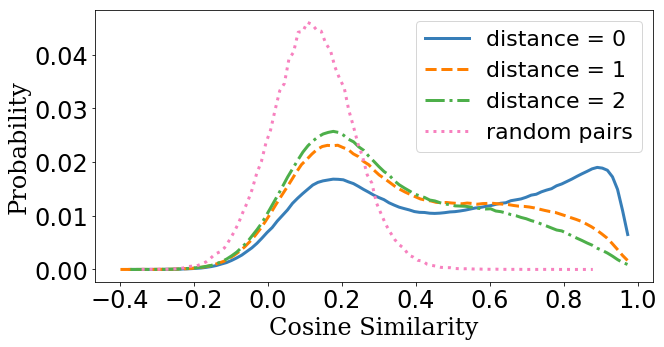}
        \caption{GEN Encoder}  \hfill
        % \label{fig:densityPlot}
    \end{subfigure} \hfill
\caption{The distributions of cosine similarities between queries in search sessions that are adjacent (1), separated by one (2) or two (3) other queries in the session, or between random sampled queries. The distance is the number of queries the user issued in between the two queries. Query pairs were randomly sampled in Bing search sessions with more than two queries.
% Similarities of random pairs are also plotted. 
\label{fig:distance}
% The x-axes mark the pairs' cosine similarity and y-axes mark their probability. \label{fig:densityPlot}
}
\end{figure*}

%% file: table/pairclasses.tex
\pdfoutput=1
\begin{table}[t]
\centering
\caption{
Spearman's rank-order correlations between representation distances and expert labels in the four intent-transit classes:
``Topic Change'' (0), ``Explore'' (1), ``Specify'' (2), and ``Paraphrase'' (3). Related includes (1-3) and Middle is (2-3). \texttt{Human} is the average performances of authors' labels.
\label{tab:catcor}
}
\begin{tabular}{l|c|c|c} \hline \hline
\textbf{Method} 
& \textbf{All (100)} & \textbf{Related (55)}
& \textbf{Middle (45)} \\ \hline

\texttt{TF-IDF BOW}
& .642
& .459
& .285
\\

\texttt{Emb BOW (q)}
& .522
& .351
& .123
\\

\texttt{BERT Encoder}
& .626
& .436
& .327
\\

\texttt{RLM+ (t)} 
& .705
& .425
& .242
\\
\texttt{USE}
& .678
& .383
& .123
\\ 
\hline
\texttt{GEN Encoder}
& .800 & .580 & .429 \\ \hline
\texttt{Human}  
& .859
& .776
& .620
\\
\hline \hline
\end{tabular}
\end{table}

%% file: table/caseStudy.tex
\pdfoutput=1
\begin{table*}[]
\centering
\caption{
Examples of Topic Change, Exploratory and Specification query reformulation pairs, labeled by our experts.
Their cosine similarities in various distributed representation space are listed. Emb is Emb BOW (q) and RLM+ is on title (t).
\label{tab:casestudy}}
\begin{tabular}{cc|c|cccc}
\hline \hline
\textbf{First Query}                                         & \textbf{Second Query}                                                               & \textbf{Label} & \bf{Emb}   & \bf{BERT} & \bf{RLM+} & \bf{GEN} \\ \hline

 ``donald norman design of everyday things''          & ``size an image in html''                           & Topic Change      & 0.80  & 0.61  & 0.62   & 0.26 \\ \hline

``facial lotion containing menthol and phenol''      & ``sarna lotion''                                     & Explore      & 0.65  & 0.67  & 0.51   & 0.61 \\

``fitness social fresno''                            & ``orange theory fitness fresno''                    & Explore      & 0.90  & 0.80  & 0.84   & 0.81 \\ \hline

``how to change autofill settings''                  & ``how to change autofill settings chrome''           & Specify      & 0.99  & 0.92  & 0.96   & 0.84 \\
``cdma''                                             & ``cdma in a cell phone''                             & Specify      & 0.53  & 0.76  & 0.50   & 0.75 \\
\hline \hline
\end{tabular}

\end{table*}

%% file: Conclusion.tex
 %\newpage
\section{Conclusion and Future Work}
\pdfoutput=1

This paper presented \texttt{GEN} \texttt{Encoder}, a neural system that learns a distributed representation space that captures user intent.
Search queries are often insufficient to express what a user wants, but user clicks provide implicit feedback signals for their search intent.
Instead of manually defining intent categories or an ontology, 
\texttt{GEN} \texttt{Encoder} utilizes user clicks as weak supervision of user intent, and learns end-to-end its encoder architecture by mapping co-click queries together in its representation space.
We further fine-tuned \texttt{GEN} \texttt{Encoder} to improve it's generality.

We used an intrinsic evaluation task -- query intent similarity modeling -- to evaluate the quality of query representations.
\texttt{GEN} \texttt{Encoder} showed significant, robust, and large margin improvements over a wide range of previous text representation methods, and its advantages thrive on the hard dataset.
A series of ablation studies illustrate: 1) the necessity of user feedback signals in learning a generic intent representation space, 2) the differences between representations learned from surrounding texts and representations learned from user clicks, and 3) the improved generalization ability from multi-task learning.

An effective distributed representation of user intent has many implications in information retrieval.
This paper demonstrates how it helps alleviate the sparsity of long tail queries with approximate nearest neighbor search in the continuous space.
Though a query may never appear before, its search intent might have been expressed by other queries in the search log.
We show that, with an index of 700 million queries' \texttt{GEN} encodings, ANN search finds such co-intent queries with high coverage, sufficient accuracy, and practical latency. 
Incorporating the observations from ANN queries significantly alleviates the sparsity of tail traffic; it reduces the fraction of unseen queries by 35\%-50\% relatively.
In practice, efficient ANN search with effective generic intent embeddings has wide impact in various components of Bing search systems.

The last experiment demonstrates the emergent behavior of \texttt{GEN} \texttt{Encoder}'s representation space, whose distances separate commonly recognized user behaviors in search sessions.
The distances of query reformulations in search sessions follow a strong bi-modal distribution, and correlate well with the four types of query reformulation behaviors: ``Topic Change'', ``Exploratory'', ``Specification'', and ``Paraphrase''.
This suggests \texttt{GEN} \texttt{Encoder} can be used to understand user's information seeking behaviors and provide more immersive search experiences.
For example, it can be used to identify sessions that will follow a trajectory such as learning a new topic or completing a specific task; which opens the opportunity for more direct answers, more engaging question suggestions, and a more conversational interaction with the search engine.

Having a meaningful generic representation for user search intents suggests many downstream applications; this paper can only demonstrate a few of them. We are delighted to make Gen Encoder available through
\url{https://aka.ms/GenEncoder} to facilitate more exploration of its potential usage.